\documentclass[a4paper]{article}

\usepackage[utf8]{inputenc} 
\usepackage{hyperref}       
\usepackage{url}            
\usepackage{amsfonts}       
\usepackage{amsmath}
\usepackage{graphicx}
\usepackage{caption}
\usepackage{float}
\usepackage[left=2cm,right=2cm,top=2cm,bottom=2cm]{geometry}
\usepackage[english]{babel}
\usepackage{amsmath}
\usepackage{graphicx}
\usepackage[colorinlistoftodos]{todonotes}
\usepackage{float}
\usepackage{dsfont}
\usepackage{bm} 
\usepackage{url}
\usepackage{booktabs} 
\usepackage{multicol}
\usepackage{multirow}
\usepackage{gensymb}
\usepackage{tabularx}
\usepackage{cite}
\usepackage[normalem]{ulem}
\DeclareCaptionFormat{sanslabel}{#3}%
\usepackage{pifont}
\usepackage{amssymb}

\begin{document}

\Large
\begin{center}
\vspace{1cm}
\textbf{Non invasive live imaging of a novel RPE stress model with Dynamic Full-Field OCT}

\vspace{0.5cm}

\normalsize
Kassandra Groux$^{1,2}$, Anna Verschueren$^{2,3}$, Céline Nanteau$^{3}$, Marilou Clémençon$^{3}$, Mathias Fink$^{1}$, José-Alain Sahel$^{2,3,4,5}$, Claude Boccara$^{1}$, Michel Paques$^{2}$, Sacha Reichman$^{3}$ and Kate Grieve$^{2,3*}$

\end{center}
\scriptsize
$^1$Institut Langevin, ESPCI Paris - PSL, CNRS, 10 rue Vauquelin, Paris, 75005, France\\
$^2$Paris Eye Imaging Group, Quinze-Vingts National Eye Hospital, INSERM-DGOS, CIC 1423, 28 rue de Charenton, Paris, 75012, France\\
$^3$Institut de la Vision, Sorbonne Université, INSERM, CNRS, 17 rue Moreau, Paris, F-75012, France\\
$^4$Department of Ophthalmology, Fondation Ophtalmologique Adolphe de Rotschild, F-75019 Paris, France\\
$^5$Department of Ophthalmology, The University of Pittsburgh School of Medicine, Pittsburgh, PA 15213, United States of America\\
$^*$kategrieve@gmail.com (Supplementary available on request)

\normalsize

\section*{Abstract}
Retinal degenerative diseases lead to the blindness of millions of people around the world. In case of age-related macular degeneration (AMD), the atrophy of retinal pigment epithelium (RPE) precedes neural dystrophy. But as crucial as understanding both  healthy  and  pathological  RPE  cell  physiology  is  for  those  diseases,  no  current  technique allows subcellular \textit{in vivo} or\textit{ in vitro} live observation of this critical cell layer. To fill this gap, we propose dynamic full-field OCT (D-FFOCT) as a candidate for live observation of \textit{in vitro} RPE phenotype. In this way, we monitored primary porcine and human stem cell-derived RPE cells in stress model conditions by performing scratch assays. In this study, we quantified wound healing parameters on the stressed RPE, and observed different cell phenotypes, displayed by the D-FFOCT signal. In order to decipher the subcellular contributions to these dynamic profiles, we performed immunohistochemistry to identify which organelles generate the signal and found mitochondria to be the main contributor to D-FFOCT contrast. Altogether, D-FFOCT appears to be an innovative method to follow degenerative disease evolution and could be an appreciated method in the future for live patient diagnostics and to direct treatment choice.

\section*{Article}
\subsection*{Introduction}
The retinal pigment epithelium (RPE) is the most external layer of the retina, placed between the photoreceptors which collect the light and the blood supply from the choroid \cite{strauss_retinal_2013}. The RPE layer is affected in degenerative diseases such as age related macular degeneration (AMD) \cite{boulton_role_2001,besch_inherited_2003,sparrow_retinal_2010}.
AMD is one of the leading causes of blindness in developed countries, affecting 170 million people worldwide [Klein et al 2011] \cite{klein_prevalence_2011}. In late form dry AMD, called geographic atrophy (GA), it may lead to expanding atrophic foci of the RPE. Little is known about the mechanisms underlying the propagation of RPE atrophy. Animal models were used to model AMD. However, the environmental conditions are less controllable than for a cell-based model, such as RPE cells \cite{forest_cellular_2015}, used to test drug therapies or the involvement of proteins in AMD. By following the behavior of RPE in culture in response to damage, we take a step towards mimicking formation of geographic atrophy (GA) lesions in AMD. Future developments will use multi-layered cell-based models, including other layers of the retina or the choroid to match \textit{in vivo} conditions.    

Dynamic Full-Field OCT (D-FFOCT) has recently been presented as a non-invasive live imaging technique suitable for the study of 2D and 3D cell cultures \cite{scholler_probing_2019,scholler_dynamic_2020}. 
Intracellular organelle movement is known to generate the D-FFOCT signal, but the precise identity of the organelles has not yet been deciphered. RPE cells have also been imaged with electron microscopy \cite{themes_cell_2017}, but this technique does not allow live-cell imaging, preventing the imaging of organelle dynamics. Live-cell imaging is possible with several imaging techniques, but certain requirements  have to be met. It is important to keep the cell cultures healthy to image them over long periods of time, meaning avoiding photo-toxicity. Imaging methods such as fluorescence imaging techniques, multiphoton microscopy, or widefield systems, have been applied to live-cell imaging \cite{jensen_overview_2013}. However, many of these techniques require the use of invasive fluorophores, preventing continued use of the sample after the imaging. Others, such as widefield systems, usually have a low lateral resolution and do not have optical sectioning for 3D imaging. The use of spinning-disc confocal microscopy for RPE live-cell imaging has been demonstrated \cite{toops_detailed_2014,rathnasamy_live-cell_2018}, but shows significant problems with autofluorescence background in images and the impossibility to reuse the samples after the imaging due again to the use of invasive fluorophores.

In this article, we propose the use of D-FFOCT to perform non invasive live imaging of RPE cell cultures. We validate a novel RPE stress model, provide a new tool to quantify RPE wound healing parameters, and identify the organelles responsible for the D-FFOCT signal.

\subsection*{Results}

\paragraph{RPE stress model and quantification methods}

\begin{figure}[H]
\centering\includegraphics[scale=1]{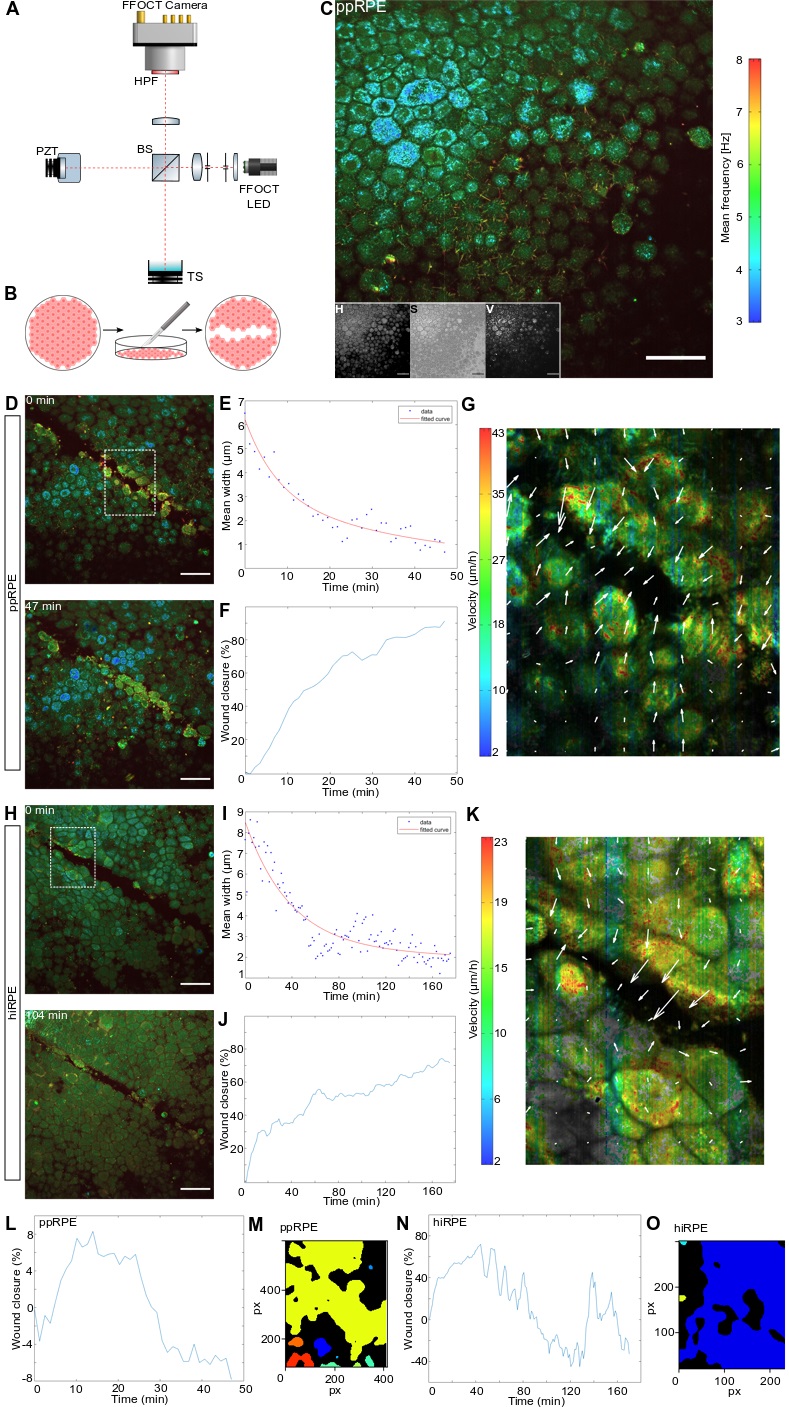}
\caption{}
\label{fig_1}
\end{figure}

Primary porcine RPE cell cultures (ppRPE) (n=5), which have the advantage of being highly pigmented as human mature RPE cells, and human induced pluripotent stem cell derived RPE (hiRPE) at an early development stage (unpigmented) \cite{reichman_generation_2017,reichman_confluent_2014} (n=9) were grown on poly-carbonate (PC) membranes.  
The setup used for D-FFOCT imaging is shown in Fig.\ref{fig_1} A. We followed the evolution of the RPE cell cultures over periods from 1 to 6 hours (3 hours on average) of D-FFOCT live imaging after inducing stress. Stress was created by performing a scratch assay on the cell layer with a scalpel blade, allowing us to generate a focal and easily reproducible stress damage to the epithelium as shown in Fig.\ref{fig_1} B (see Materials and methods). Fig.\ref{fig_1} C shows a D-FFOCT image of an intact ppRPE sample with the corresponding colorbar, where the three channels H, S and V of image processing (see Materials and methods) are represented in the lower left corner. Sets of parallel cultures were imaged at fixed timepoints with immunohistochemistry (an histology technique using bio-engineered antibodies coupled with fluorescent probes to stain specific proteins and cellular structures within tissues), in order to identify the organelles generating the D-FFOCT signal. 

Depending on the origin of the cells (ppRPE or hiRPE) and the width of the scratch, cell behaviours differed. Two main features are usually used to quantify wound healing: the speed of closure, i.e. the evolution of the width of the scratch, and the wound closure, measuring the evolution of the area of the scratch \cite{abu_khamidakh_wound_2018}. Semi-automatic segmentation based software, named Scratch Assay Velocity Evolution (SAVE) Profiler, (see Materials and Methods), was developed to segment the wound and calculate the width and area of the scratch in order to quantify wound closure. Our SAVE Profiler method outperformed the reference methods using optical flow calculations \cite{barron_performance_1992} and the Cell Profiler using the Wound Healing example \cite{lamprecht_cellprofiler_2007}.

\paragraph{RPE behaviour observed after stress}

In this RPE stress model, we separated three cases: scratches inferior to 25$\mu m$ wide (Fig.\ref{fig_1}), scratches between 25 and 100$\mu m$ (Fig.\ref{fig_2}, and scratches superior to 100$\mu m$ (Fig. supplementary 1).

Figure \ref{fig_1} shows wound closure on both ppRPE (\ref{fig_1}~D to J) and hiRPE (\ref{fig_1}~H to K) cell cultures for scratches smaller than 25$\mu$m. In a ppRPE sample, after 47 min, the wound was completely closed (Fig.\ref{fig_1}~D) (see the whole acquisition in Movie S2). Calculations of scratch assay evolution were made over a small area (white dotted square on Fig.\ref{fig_1}~D) to reduce calculation time. Fig.\ref{fig_1}~F plots the wound closure, which represents the evolution of the wound area, as calculated with the SAVE profiler. A quasi complete closing was achieved, with the wound closure reaching over 90$\%$ during the acquisition. Fig.\ref{fig_1}~E represents the evolution of the mean scratch width (blue points). This evolution could be fitted with a double exponential $f(x)=a\times exp(b\times x)+c\times exp(d\times x)$. The characteristic time of closing obtained with this fit was approximately 18 minutes. The closest distance of closing of around 1$\mu m$ was reached after 23 minutes, corresponding to the average interstice between cells which is not affected by the scratch. Average wound closing speed for this ppRPE sample was thus calculated to be 15.7$\mu m/h$. Fig.\ref{fig_1}~G presents an optical flow calculation over the same area. The velocity of movements throughout the acquisition is shown, with two fronts of motion in opposite directions revealed by the arrows. By averaging the velocities (i.e. the absolute values without taking the direction into account) plotted in Fig.\ref{fig_1}~G, the medium speed of one front can be calculated as 8.25$\mu m/h$. As there are two opposite moving fronts, the total velocity of closing is the multiplication of the average speed of one front by 2, giving a total speed of 16.5$\mu m/h$. This result is consistent with the results obtained with our custom-written SAVE profiler software. The same analysis was performed on hiRPE cell cultures (Fig.\ref{fig_1}~H to K) (see the whole acquisition in Movie S3). Fig.\ref{fig_1}~H shows the cell layer, almost closed, 103 minutes after the first image was acquired. The white dotted square on Fig.\ref{fig_1}~H represents the area over which calculations were performed. The wound does not completely close (Fig.\ref{fig_1}~H), reaching an average wound closure of 70$\%$ (Fig.\ref{fig_1}~J). As for the ppRPE scratch assay, the mean scratch width was plotted in Fig.\ref{fig_1}~I, where it could also be plotted with a double exponential. In this case, the characteristic time of closing was around 55 minutes, and an average limit of 2.5$\mu m$ was reached after approximately 60 minutes. The average speed of closing was therefore 6.5$\mu m/h$. The results of the optical flow method are shown in Fig.\ref{fig_1}~K, clearly showing two borders which progress in two different directions, but the upper border is more active than the lower (see the aspect of arrows in Fig.\ref{fig_1}~K). The average speed calculated on the optical flow representation was 4.05$\mu m/h$, or multiplying this result by 2 to be cohesive with two different borders, we obtain 8.1$\mu m/h$, consistent with the results of the SAVE profiler. The SAVE Profiler was validated on the Wound Healing example of Cell Profiler. Cell Profiler showed less effective segmentation (Fig.\ref{fig_1}~M for the ppRPE sample and O for the hiRPE sample) as it does not uniquely target the wound. The Cell profiler results of wound closure (in Fig.\ref{fig_1}~L and N) are therefore not consistent with our observations. Moreover, SAVE profiler was twice as fast as Cell Profiler.

As the scratch assays were performed manually with a scalpel blade, the initial width of the wound could vary. In wounds sized between 25$\mu m$ and 100$\mu m$, the wound failed to close. Fig.\ref{fig_2} presents failure of wound closure on wounds of greater than 25$\mu m$ width in ppRPE and hiRPE. In ppRPE ((Fig.\ref{fig_2}~A) (see the whole acquisition in Movie S4), the first and the last images of the acquisition are 665 minutes apart. Calculations were performed over the area in the white dotted square. The evolution of the scratch width (Fig.\ref{fig_2}~B) and the wound closure (Fig.\ref{fig_2}~C) show that the cell layer first tends towards closure for the first 50 minutes, followed by a period of retraction. Fig.\ref{fig_2}~D presents the results of the optic flow calculations: the arrows show an initial movement towards wound closure, but the inside of the wound remains stationary, in contrast to wounds under 25 $\mu m$ size in Fig.\ref{fig_1}. hiRPE generally showed similar behaviour (Fig.\ref{fig_2}~E-H) (see the whole acquisition in Movie S5). Two cells appear with a higher velocity: these cells were dying and detached from the border of the wound. Cell Profiler was more effective on non-closing than on closing wounds and gave consistent results with those calculated with SAVE Profiler (see Fig.supplementary 2).

In summary, small wounds (under 25$\mu m$) lead to a repair of the damaged cell layer, while large wounds (between 25 and 100 $\mu m$) tend to create a movement towards closure followed by a retraction. For wounds larger than 100$\mu m$, we observed a direct expansion of the cell layer with no evidence of attempt at closure as shown in Fig.supplementary 1. Comparing repair processes for the two types of RPE cells, ppRPE tend to close faster, with a mean speed between 15 and 18 $\mu$m/h, while hiRPE close with a speed between 3 and 8 $\mu$m/h. Moreover, ppRPE tend to reach a wound closure of almost 100$\%$, while for hiRPE this is not always the case. The statistics of the scratch assays performed are compiled in supplementary Tab.supplementary 2.

\paragraph{Validation of D-FFOCT signal with immunohistochemistry}

In D-FFOCT images, the dynamic profile of the cells around the border of the scratch evolve throughout the acquisition. In order to understand which organelles were involved in these phenomena, ppRPE and hiRPE scratch assays and wound healing observed over a period of several hours with D-FFOCT were paralleled by a set of identical cultures fixed at different timepoints for immunohistochemistry validation, via fluorescent labeling of different organelles. 
Actin filaments (labelled using a high-affinity F-actin probe conjugated to fluorescent dye, Fig.\ref{fig_3}~A) contribute to the cytoskeleton of the cell in varied cell structures in RPE: the cell cortex (first column of Fig.\ref{fig_3} A), the microvilli (second column of Fig.\ref{fig_3} A) and the fillipods (last column of Fig.\ref{fig_3} A). The cytoskeleton includes the cortex sustaining the plasmic membrane, which is static, while the microvilli and the fillipods are both very active. As the cortex is static, it appears black on D-FFOCT images, while microvilli and fillipods were very active: moving rapidly and constantly (high intensity and red) in D-FFOCT. Individual microvilli were easily visible at the surface of RPE cells on D-FFOCT images compared to immunohistochemistry images, and could be imaged in 3D by performing a stack acquisition in depth. Their movement could be followed through live image sequences using a custom GPU-computing software (Holovibes \url{http://holovibes.com/} \cite{atlan_holovibes:_2014} (see Movie S7)), for the first time to our knowledge. After a closed scratch assay, an accumulation of actin is visible in immunohistochemistry (Fig.\ref{fig_4} D). It is composed of the border cells of the wound which seem to disintegrate after the closing, as observable in the corresponding D-FFOCT image. This is coherent with the literature \cite{geiger_transmembrane_2001}.

Pigment particles (ovular, 2 - 3 $\mu m$ x 0.5 - 1 $\mu m$) absorb light, making them visible in conventional microscopy and phase contrast microscopy (here we used Normanski interference contrast and superimposed the fluorescent phalloidin image corresponding in green as reference image). They were individually resolved throughout the D-FFOCT images in Fig.\ref{fig_3}~B. As primary cultures, ppRPE were highly pigmented, while hiRPE contained little pigment. Pigment was active but very slow moving, giving a bright blue dynamic profile in D-FFOCT. The ovular shapes measured around 3 $\mu m$ x 1 $\mu m$, corresponding to the literature \cite{themes_cell_2017}. The difference in pigment signal between ppRPE and hiRPE is highly visible (Fig.\ref{fig_3} B). This assumption was confirmed by phase contrast microscopy of the samples, showing a large difference in melanin concentration.

The dynamic profile was different for cells far from the wound and for cells close to the wound (Fig.\ref{fig_3} E). The main parameter that changes in the dynamic profile was the colour, meaning the frequency of sub-cellular movements. 

Golgi apparatus (cyan label) showed a uniform distribution and activity across all cells, regardless of cell damage or wounding in Fig.\ref{fig_3}~C. Golgi therefore contributed evenly to the D-FFOCT signal throughout the sample. 
Lysosomes (magenta label in Fig.\ref{fig_3}~C and D,  using two different antigenes for confirmation), vesicles that contain hydrolytic enzyme and act as the waste disposal system of the cell, were visible in a small proportion of cells which detached from the culture and became mobile. Their high activity and mobility during late apoptosis caused them to have a bright red dynamic profile in D-FFOCT, and they tended to migrate toward the wound during closure.
In cells neighboring the wound, mitochondria (yellow label in Fig.\ref{fig_3} C and green label in Fig.\ref{fig_3} D, using two different antigenes for confirmation) change aspect in the cells circled in yellow on stressed cell images. Therefore, we can affirm that mitochondria are involved in D-FFOCT signal.
To go further in this observation, we compared immunohistochemistry imaging of mitochondria to D-FFOCT images at four different timepoints (0 hour, 1 hour, 2 hours and 3 hours) on both normal cells and stressed cells in Fig.\ref{fig_4} A. While the mitochondria organisation in immunohistochemistry and their dynamic profile in D-FFOCT were identical over time for normal cells, we observed that mitochondria change form from a filament network to isolated round spheroids in stressed cells (two right columns), which is coherent with the literature \cite{arnoult_mitochondrial_2007,ahmad_computational_2013,miyazono_uncoupled_2018}. These changes were visible in D-FFOCT thanks to their form, their high activity revealed by their colour changing from green to red, and their enhanced brightness. As the resolution of D-FFOCT is 0.5$\mu m$, healthy individual mitochondria (diameter 0.4$\mu m$) were not resolved but gave a general background signal, while the bright dots in stressed cells suggest that they are damaged mitochondria which formed larger spheres that could be individually distinguished. \cite{ahmad_computational_2013,miyazono_uncoupled_2018}
Dying cells are distinguishable in D-FFOCT. Cells undergoing apoptosis or necrosis were labelled with propidium iodide (magenta label) in Fig.\ref{fig_4} B. These cells (circled in yellow) show a condensed nucleus and a change in mitochondria organisation (green label), thus undergo apoptosis. In D-FFOCT, these cells appear with a faster subcellular activity (red dynamic profile) and a condensed nucleus with also a higher activity. The size of nuclei were measured and compared in both imaging techniques and the results were consistent (diameter under 7$\mu m$ for condensed nuclei).
However D-FFOCT signal is not related to dead cells: those few cells which died (showing a high concentration of lysosomes in magenta label) following apoptosis and remained mobile in the cultures are dark in D-FFOCT due to absence of organelle activity (Fig.\ref{fig_4} C), but the structure is visible in static FFOCT, recorded in parallel, which reveals static structural rather than functional information. 

Our results are confirmed by those shown in electron microscopy \cite{themes_cell_2017}, where the RPE cytoplasm is mainly filled with mitochondria (usually arranged in a network for unstressed cells) and pigment granules (of similar size to those measured here with D-FFOCT). Electron microscopy of microvilli is also consistent with what we observed with D-FFOCT, but is not capable of showing the movements that were visible with D-FFOCT.

\subsection*{Discussion}
D-FFOCT is able to non invasively show mitochondrial dynamics, microvilli and fillipods, and pigment distribution in RPE cells via live imaging in real time. Comparison of wound healing parameters quantified using optical flow \cite{barron_performance_1992}, semi-automatic segmentation methods (SAVE Profiler) and Cell Profiler software \cite{lamprecht_cellprofiler_2007}, showed that our custom-developed SAVE Profiler succesfully combines results from both optical flow and Cell Profiler in a shorter calculation time. 
Wound healing parameters such as speed agreed with the literature \cite{hergott_inhibition_1993}. In our experiments (between 1 hour and 6 hours), we only saw migration of cells especially for small wounds, which is coherent with Hergott et al., 1993 \cite{hergott_inhibition_1993}, while large wounds tend to involve proliferation. Moreover, we saw that the integrity of the membrane was important for the closing of the wound as shown in Geiger et al., 2001 \cite{geiger_transmembrane_2001}. Jacinto et al., 2001 \cite{jacinto_mechanisms_2001} and Farooqui et al., 2005 \cite{farooqui_multiple_2005} which show that several rows of cells participate in the closing of the wound. This phenomenon was observable on closing wounds in D-FFOCT. Filipods seem to also be involved in the closing of the wound, which is confirmed by Jacinto et al., 2001 \cite{jacinto_mechanisms_2001}.
The different methods presented to calculate the evolution of scratch assays present consistent results on the speed of either closing or expansion. Differences were observed depending on the origin and maturation stage of the sample, in accordance with the literature \cite{abu_khamidakh_wound_2018}. The experiments were performed multiple times (n=5 ppRPE; n=9 hiRPE) giving the same results each time on speed and wound closure.
In small ($<$25$\mu m$) scratches, wounds closed at 16$\mu m/h$ for ppRPE and 5$\mu m/h$ for hiRPE on average. After wound closure, the actin remained thickened around the wound, some damaged mitochondria remained encompassed by other cells, and cells along the wound border remained raised above the rest of the culture surface. In wounds with large scratches ($>$25$\mu m$), a different behaviour was observed: an initial attempt at wound closure (cell sheet moving inwards) was overcome by a retractive movement away from the wound, with sliding of the whole cell sheet. This behaviour displays some similar characteristics to those seen \textit{in vivo} over long (several months to years) time periods on patients with GA lesions forming \cite{paques_adaptive_2018}.
The dynamic profile of cells was also consistent throughout the acquisitions: cells on the border of the wound show a faster (i.e. colour tending to yellow/red) and higher (i.e. brighter) signal than cells far from the wound. These observations on D-FFOCT images were consistent with the results obtained in immunohistochemistry, showing that mitochondria have a different behaviour in the damaged cells, and appears to be the organelles undergoing the most drastic phenotype change.
In order to get closer to human \textit{in vivo} RPE cells, we tried to study hiRPE at a mature development stage (pigmented, 90 days of growing). Unfortunately, we were not able to determine if the cell movements were due to the scratch assay or breaks in the cell layer caused by lack of membrane adhesion. This problem was also observed by Abu Khamidakh et al., 2018 \cite{abu_khamidakh_wound_2018}.
Future developments of this study would be to automatize the realization of scratch assays with a new setup combining a laser to cut the cell layer with better repeatability. 
We could also test the addition of molecules to change the speed of closure \cite{croze_rock_2016}.
We showed that D-FFOCT combined with custom-developed calculation software allow imaging and study of degeneration in RPE cell cultures, which may contribute to the comprehension and the following of \textit{in vivo} degenerative retinal diseases (such as AMD) evolution, and could be an appreciated method for live patient diagnostics and direct treatment choice in the future. As a marker of mitochondrial dynamics and fragmentation and thus of their activity \cite{miyazono_uncoupled_2018}, D-FFOCT may also be used in the understanding of optic nerve disease, where mitochondria are implicated. Parallels of wound healing or wound retraction in the RPE layer are found with \textit{in vivo} adaptive optics imaging of AMD patients over year long periods \cite{gocho_adaptive_2013}.

\addtocounter{figure}{-1}
\begin{figure}
  \caption{\small \textbf{D-FFOCT imaging \& scratch techniques (A to C) and results of closing ($<$25$\mu m$) scratch assays (D to O). (scale-bar: 50 $\mu m$)} \textbf{A} Drawing of our custom-built Full-Field OCT system. BS: Beam-Splitter; HPF: High-Pass Filter; PZT: Piezo-electric Translation; TS: Translation stage (for the sample). \textbf{B} Schematic of the RPE cell layer before and after the scratch assay (from left to right).  \textbf{C} Recombined three channels of HSV computation and the three different channels in lower left. The colorbar represents the frequency variations of the sample (Hue channel). \textbf{D to G} Results of the analysis of a closing scratch assay on a primary porcine RPE cell culture (ppRPE). \textbf{D} Beginning and end (i.e. closing) of the imaging of the scratch assay. The dotted white square corresponds to the area used for calculations. \textbf{E \& F} Plots of the evolution of scratch width and wound closure over the acquisition, calculated with SAVE Profiler. \textbf{G} Optical flow calculations showing velocity and motion direction. \textbf{H to K} Results of the analysis of a closing scratch assay on a hiRPE cell culture. \textbf{H} Beginning and end (i.e. closing) of the imaging of the scratch assay. \textbf{I \& J} Plots of the evolution of scratch width and wound closure over the acquisition, calculated with SAVE Profiler. \textbf{K} Optical flow calculations. \textbf{L \& N} Wound closure calculated on the closing ppRPE and hiRPE scratch assays with Cell Profiler. \textbf{M \& N} Segmentations of the final images of the acquisitions with Cell Profiler, showing the superior performance of SAVE Profiler.}
\end{figure}

\begin{figure}[H]
\centering\includegraphics[scale=0.7]{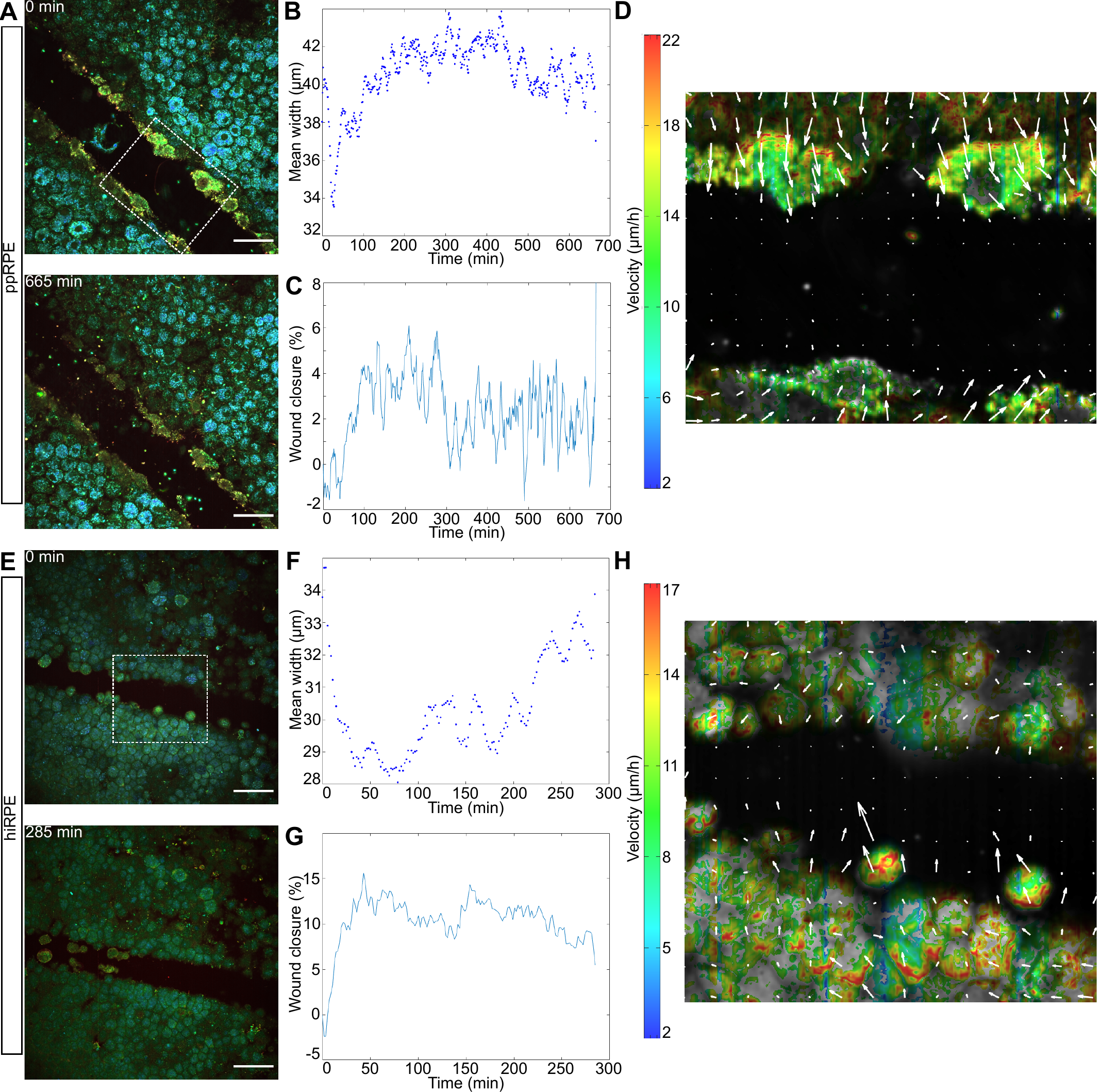}
\caption{\small \textbf{Results on scratch assays ($>$25$\mu m$) on ppRPE and hiRPE cell cultures, failing to close entirely.} \textbf{A} Beginning and end of the imaging of a scratch assay on ppRPE cell culture. \textbf{B \& C} Evolution of scratch width and wound closure calculated with our program and \textbf{D} optical flow calculations. \textbf{E} Beginning and end of the imaging of a scratch assay on hiRPE cell culture. \textbf{F \& G} Evolution of the scratch width and wound closure, calculated with the SAVE profiler and \textbf{H} optical flow calculations. (Scale-bar: 50 $\mu m$)}
\label{fig_2}
\end{figure}

\begin{figure}[H]
\centering\includegraphics[scale=0.8]{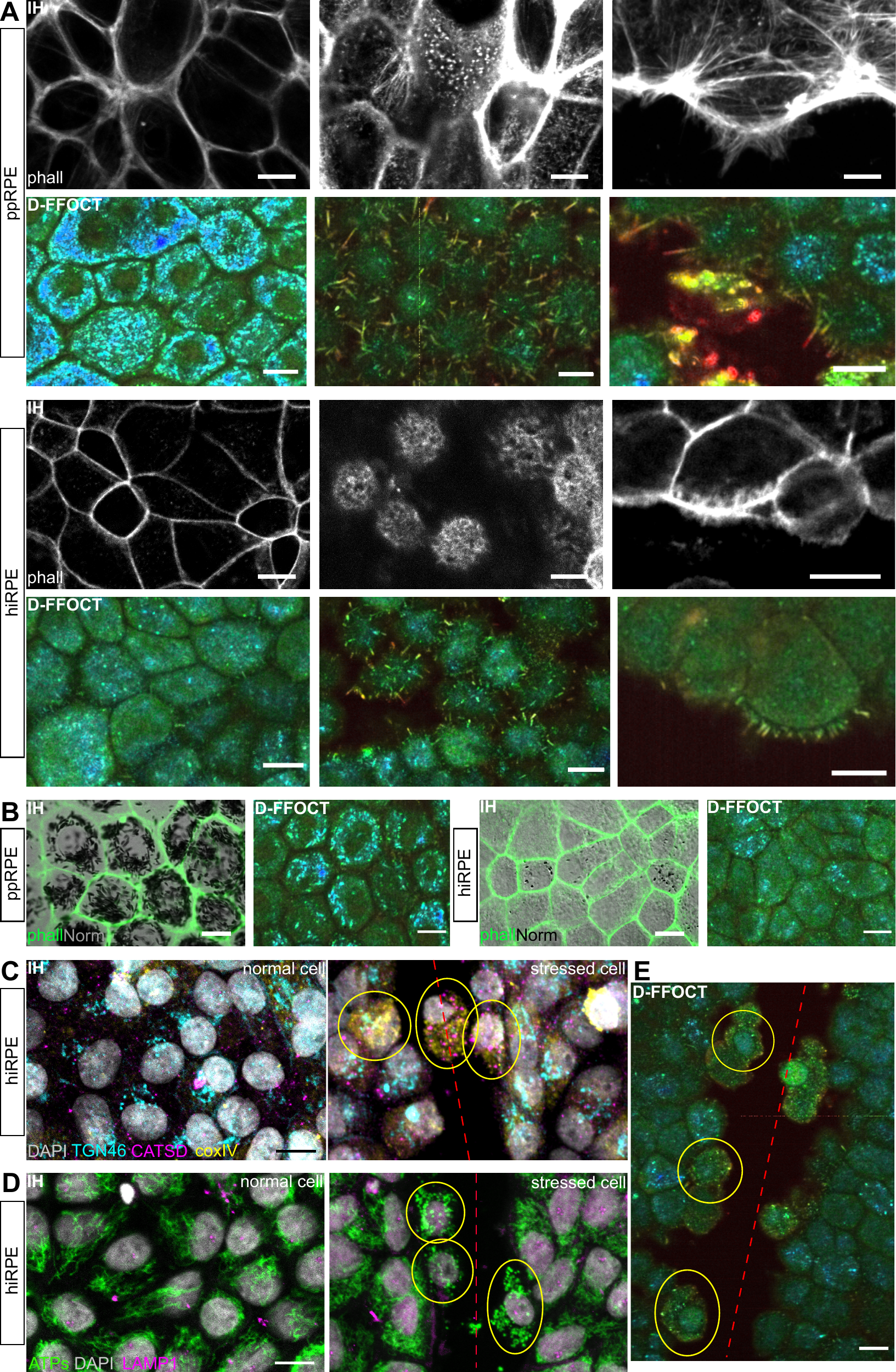}
\caption{\small \textbf{Immunohistochemistry validation of organelles.} Nuclei were labelled with DAPI. Red dashed lines show the scratch direction. \textbf{A} Actin filaments (phalloidine in IH) validation on ppRPE and hiRPE. The left column shows the cell cortex (static) appearing dark in D-FFOCT. The center column presents microvilli. Their visualisation is easier in D-FFOCT: they appear as red cillias on top of RPE cells. The right column shows filipods on the side of the cells close to the wound, appearing in red in D-FFOCT such as microvilli. \textbf{B} Comparison of pigment granules on both ppRPE (highly pigmented) and hiRPE (few pigments). Actin filaments (cytoskeleton) are labelled with phalloidine in IH, while pigments were imaged with phase contrast imaging. \textbf{C} Golgi apparatus (TGN46 in cyan), lysosomes (CATSD in magenta) and mitochondria (coxIV in yellow) IH comparison between normal cell and stressed cell (circled in yellow). \textbf{D} Mitochondria (ATPs) and lysosome (LAMP1) validation in IH between normal cell and stressed cell (circled in yellow). \textbf{E} Related D-FFOCT image for \textbf{C - D} showing both normal cells and stressed cells (circled in yellow). IH: immunohistochemistry. (Scale-bar: 10$\mu m$)}
\label{fig_3}
\end{figure}

\begin{figure}[H]
\centering\includegraphics[scale=0.8]{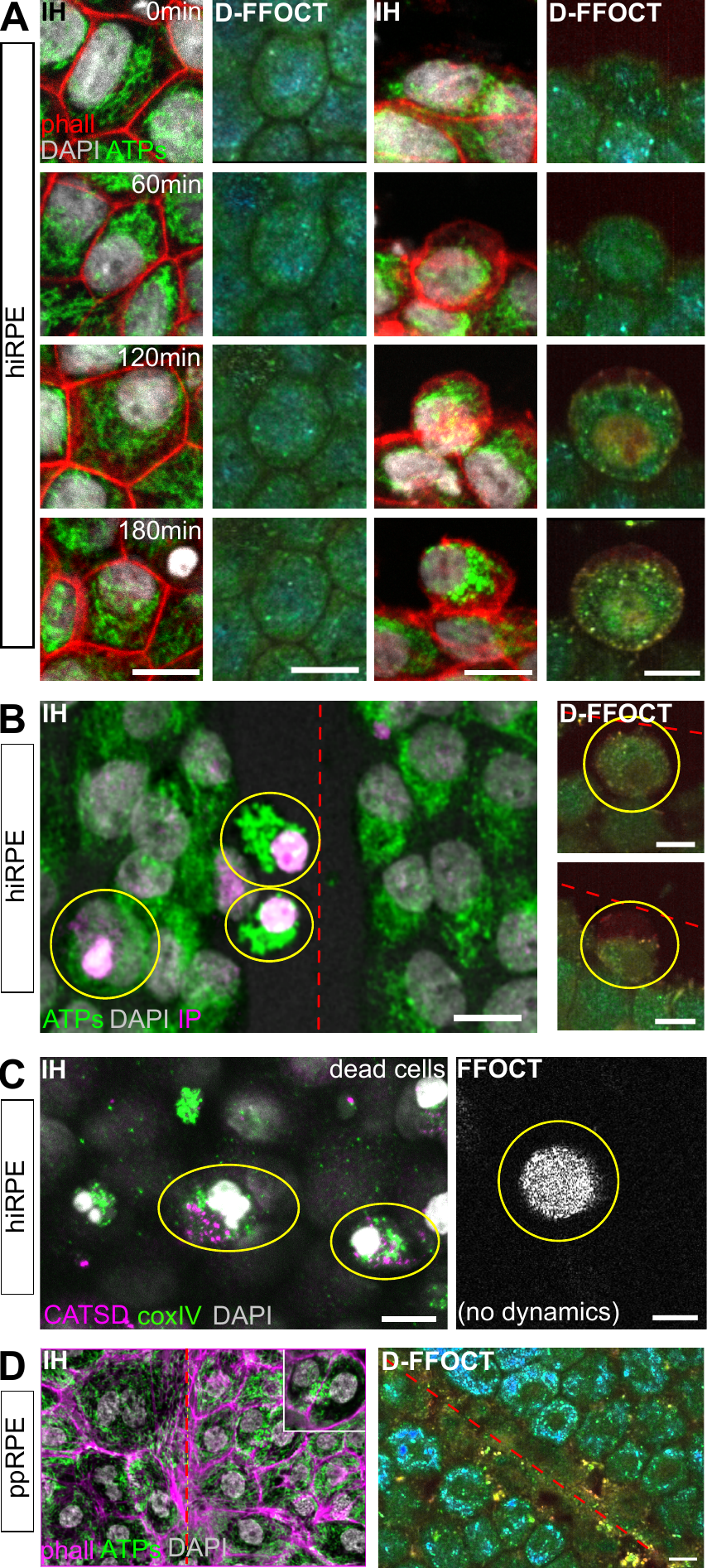}
\caption{\small \textbf{Cell phenotypes.} Nuclei were labelled with DAPI. Red dashed lines show the scratch direction. \textbf{A} Mitochondria phenotype evolution over 3 hours. The two left columns show the evolution of a normal cell, with a constant phenotype. The two right columns present the evolution of a stressed cell, with changes in mitochondria aspect (IH) and in dynamic profile (D-FFOCT). \textbf{B} Dying cells (circled in yellow) 2 hours after the scratch, labelled with propidium iodide (IP in magenta), showing a condensed nuclei. Mitochondria are labelled with ATPs (green). Corresponding cells in D-FFOCT on the right. \textbf{C} Dead cells (circled in yellow) floating over the cell culture. Lysosomes labelled with CATSD (magenta), mitochondria with coxIV (green) in IH. Corresponding cells do not exhibit any dynamic signal and are only visible in static (structural not functional) FFOCT. \textbf{D} Imaging of a closed scratch assay on a ppRPE sample. Actin filaments labelled with phalloidine (magenta) and mitochondria with ATPs (green) in IH. Corresponding D-FFOCT imaging. IH: immunohistochemistry. (Scale-bar: 10$\mu m$)}
\label{fig_4}
\end{figure}

\newpage
\section*{Materials and methods}

\subsection*{Dynamic Full-Field OCT: Experimental setups and image processing}

Optical Coherence Tomography (OCT) is an imaging technique which allows non invasive scanning of a sample in depth, invented in 1991 \cite{huang_optical_1991}. Time-domain Full-Field OCT (FFOCT) \cite{beaurepaire_full-field_1998,dubois_high-resolution_2002,dubois_ultrahigh-resolution_2004} is an interferometric imaging technique which is an \textit{en face} variant of OCT. This configuration, based on a Linnik interferometer i.e. an OCT setup with microscope objectives in the reference and sample arms, allows the recording of 2D images on a CMOS camera in a single shot. By scanning the sample in depth, 3D volumes of the structure of the sample can be acquired (see Movie S1). For our study, we used a laboratory-designed time-domain FFOCT, see Fig.\ref{fig_1}~A. This setup is composed of a water-immersion microscope objective in each arm (Nikon NIR APO 40x 0.8 NA), giving a lateral resolution of 0.5$\mu m$ for a total field-of-view of 320$\times$320$\mu m^2$. The axial resolution of 1.7$\mu m$ is also determined by the microscope objectives, due to their high numerical aperture. The camera used is an Adimec camera (Quartz 2A750, Adimec), custom-built for our purpose. The illumination is performed by an LED, centered at 660$nm$ (M660L3, Thorlabs), which is separated into reference and sample arms by a beamsplitter (BS028, Thorlabs). The reference arm is composed of a silicon mirror (to approach a reflectivity match with the biological samples), affixed to a piezo-electric translation stage. 
The piezo-electric translation stage in the reference arm is used to generate a phase shift. A pair of $\pi$-phase shifted images are recorded and subtracted to extract the coherent part of the interference signal between the imaged sample plane and the reference mirror.

Recently, we showed that a new type of contrast could be extracted from images acquired with FFOCT \cite{apelian_dynamic_2016}. By acquiring several hundred images without using the piezo-electric translation of the reference arm,  and calculating the standard deviation over the image stacks, we are able to extract the intrinsic motion of the biological sample. The intrinsic motion is created by the movements of the organelles inside the cells constituting the biological sample \cite{thouvenin_cell_2017}. These dynamics are referred to as the "dynamic profile" throughout this article.
The image computation of the dynamic profile is based on a power spectrum analysis, as shown in Scholler et al \cite{scholler_dynamic_2020}. For this purpose, 512 images are recorded at 100Hz on our FFOCT setup. The power spectrum analysis is then performed on each voxel of the stack of images. The study of the time variations over each voxel helps to highlight the intra-cellular motion recorded. These variations are coded in the Hue-Saturation-Value colorspace, which is an orthogonal colorspace, providing us three different channels to compute different physical parameters. 

The mean frequency of the recorded intra-cellular motion codes for the Hue channel, which represents the colour in the image. The colour ranges from blue, representing low frequencies, to red, coding for high frequencies. 
The saturation channel is coded by the inverse of the frequency bandwidth of each voxel. For a broad bandwidth, meaning there is a large range of frequencies, the saturation will be low, creating a greyish appearance. On the contrary, for a sharp bandwidth, where a specific frequency is emphasised, the saturation will be high, creating a vivid colour.
Finally, the value, which codes for the intensity in the image, is calculated as the standard deviation over a moving window of 50 images, which are then averaged, to give the final intensity highlighting the intra-cellular motion.
Finally, the three channels are combined to create a coloured image, representing the dynamic profile of the imaged sample, see Fig.\ref{fig_1}~C.

\subsection*{RPE cell cultures}

RPE is a cell monolayer composed of hexagonal cells with a basal nucleus, containing a variety of organelles. On the basal surface, the RPE is linked to the choroidal vasculature, while on the apical surface, RPE cells have microvilli which grab the photoreceptor outer segments to maintain the integrity of the retina \cite{bonilha_retinal_2006}. Different methods have been developed to grow RPE \cite{fronk_methods_2016}. We described here the methods used for both ppRPE and hiRPE samples. 

Porcine eyes were bought at a local slaughterhouse (Guy Harang, Houdan, France) in agreement with the local regulatory department and the slaughterhouse veterinarians (agreement FR75105131). This procedure adheres to the European initiative for restricting animal experimentation as not a single animal was killed for our experimentation. Eyes were taken from animals sacrificed daily for human consumption. Eyes were cleaned from muscle, and incubated for 4 minutes in Pursept-AXpress (Merz Hygiene GmbH, Frankfurt, Germany) for disinfection. The anterior portion was cut along the limbus to remove the cornea, lens and retina. A solution containing 0.25$\%$ trypsin-EDTA (Life Technologies, Carlsbad, CA, USA) was introduced for 1 hour at 37$^{\circ}$C in the eyecup. RPE cells were then gently detached from the Bruch's membrane and resuspended in Dulbecco's Modified Eagle medium (DMEM, Life Technologies) supplemented with 20$\%$ Fetal Bovine Serum (FBS, Life Technologies) and 10 $\mu$g/ml gentamycin (Life Technologies). Purified cells from one eye were pooled and plated in 2 Transwell inserts (reference: 3412 Corning). Cells were allowed to grow in an incubator with a controlled atmosphere at 5$\%$ CO$_{2}$ and 37$^{\circ}$C. The culture medium was renewed 24 hours after the first seeding.

HiRPE were generated using established protocol using AHF1pi2 hiPSC clone as described in Reichman et al., 2017 \cite{reichman_generation_2017}. Thawed hiRPE cells at passage 1 (hiRPEp1) were seeded on Geltrex (ThermoFisher) precoted flask at 50.000 cells/cm$^{2}$ and expanded in the ProN2 medium composed of DMEM/F12, 1\% MEM nonessential amino acids, 1\% CTS N2 supplement, 10 units per ml Penicillin, and 10 mg/ml Streptomycin; and the medium was changed every 2-3 days. At confluence, hiRPEp2 cells were dissociated using trypsin and replated at 100.000 cells/cm$^{2}$ on Gletrex precoted 6 well plate (reference: 3412 Corning) for D-FFOCT experiments and in p24 on Gletrex precoted glass inserts for immunostaining. All experiments were done with confluent hiRPEp3. Cells were allowed to grow in an incubator with a controlled atmosphere at 5\% CO$_{2}$ and 37$^{\circ}$C.

\subsection*{Choice of sample holder for D-FFOCT imaging and scratch assays}

In order to model \textit{in vitro} degeneration on a RPE cell culture, scratch assays were performed manually with a scalpel blade through the cell layer on the sample holder. The wounds performed ranged from 20$\mu$m wide to over 300$\mu$m.

Different materials were tested as sample holder for the cells (see Tab.supplementary 1 in Supplementary materials). The material needs to have three main characteristics: i) the cells should grow easily on it, ii) it should have a refractive index close to that of water, in order to avoid fringe artefacts caused by our interferometric technique, and iii) it should be resistant to scratches made by the scalpel blade. Most commonly, cells are grown in Petri dishes made of polystyrene, where they easily and rapidly grow. However, the refractive index of polystyrene is 1.59, far from the refractive index of water, creating fringe artefacts on the D-FFOCT images. Polytetrafluoroethylene (PTFE) membranes with a refractive index of 1.31 do not create artefacts, but PTFE is very fragile and cells do not grow easily on it. The best compromise was found to be polycarbonate (PC) membranes. These membranes have a refractive index of 1.58 but, as they are porous, the water can enter the membrane and artificially reduce the effective refractive index. Moreover, this material is quite resistant to scratches and cells grow more easily on it than on PTFE.

\subsection*{Immunochemistry preparation and imaging}

To allow both immunohistochemistry and observation of cells using confocal microscopy, parallel sets of hiRPE and ppRPE cells were cultured on glass inserts (as polycarbonate membranes do not permit confocal imaging of the epithelium). To obtain samples similar to D-FFOCT observations, scratch assays were performed on hiRPE and ppRPE cells, followed by tissue fixation using paraformaldheyde 4\% at various timepoints: 0 min (fixation just after performing the scratch), 60 min, 120 min, 180 min and 24 h. 
Propidium iodide staining was performed using a pre-fixation incubation of the samples for one hour at 37$\circ$C before cell fixation.
Immunostaining was performed using the following solution:  PBSGT, 1 x PBS containing 0.2\% gelatin: 24350262, Prolabo, and 0.5\% Triton X-100 T8787, Sigma Aldrich. A first incubation of the tissues with PBSGT alone for 2h (at room temperature, with shaking at 70 rpm) allowed blockage of non-specific binding and permeabilization. The cultures were then  incubated with the primary and secondary antibodies, in the same PBSGT added with corresponding antibodies, at 4 $\circ$C overnight  for the primary antibody and the secondary antibody.

Different organelles were labelled using the following primary antibodies:
\begin{itemize}
    \item Mitochondria: ATP synthase Subunit Beta Mouse Monoclonal Antibody (A21351 Life technologies) 1/500 and mCoxIV (mouse monoclonal [20E8] ab14744 Abcam) 1/250
    \item Nuclei: Hoescht 1/1000
    \item Actin filaments: phalloidin 1/40 (10634053 Fisher Scientific)
    \item Lysosomes: LAMP1 (ab24170 Abcam) 1/500 and CATSD (sc6486 Santa Cruz) 1/150
    \item Golgi apparatus: TGN46 (rabbit polyclonal ab50595 Abcam) 1/250
    \item Dying cells: propidium iodide (P4170, Sigma Aldrich) 1/1000
\end{itemize}
Fluorescent Secondary antibodies were produced in donkey, against rabbit, goat, and mice and coupled with Alexa 488 and 594 (1/500, Sigma aldrich).

After immunostaining, samples were then mounted in Vectashield (H1000, Vector Laboratories). 
All images were acquired with a confocal microscope, with an oil immersion objective (classical imaging) using either confocal fluorescent imaging or Normansky phase contrast imaging.
Images were observed and processed with FIJI \cite{schindelin_fiji_2012}.

\subsection*{Optical flow}

Optical flow is a method to study the motion between frames of a video. It relies on changes in the brightness pattern throughout an acquisition, and is used in navigation control and robotics, or Artificial Intelligence for example. 
Here, we first take the intensity channel of the images of an acquisition and apply a median filter to remove noise (e.g. line noise from the camera). We average 8 by 8 frames to improve the intensity and reduce the time of calculation. Then, we apply the Optical Flow from Matlab \cite{scholler_probing_2019}, using the Horn-Schunck method \cite{barron_performance_1992}. The Horn-Schunck method is based on the derivatives of the brightness of the frames, assuming there is a certain smoothness in the flow between the images. The optical flow gives magnitude, orientation, and the velocity components on x and y axes.

For the representation of the velocity and the angle of the optical flow, we choose a limit of 2$\mu$m/h as a minimum velocity. The angle is plotted by summing all the different angles calculated through the optical flow process. Velocity is plotted from the magnitude using the quiver function from Matlab, which gives a representation of the velocity with arrows. 

\subsection*{The SAVE Profiler: Custom-developed software to segment and analyze wound closure}

In order to evaluate the movement of the cells following the scratch, we wished to study the speed and the percentage of closure of the wound by directly measuring the size of the wound over a time-stack of images. For this purpose, custom software was developed in Matlab, which we name the Scratch Assay Velocity Evolution (SAVE) profiler.

The first step is to segment the scratch to create a binary image of the scratch and the cells. In order to obtain the best binary image (i.e. which considers the cell interstices, which are not associated with the scratch, as part of the cells), we applied a multiple threshold to the image and retained only the first level, which separated the scratch from the rest of the image. We performed this multiple thresholding on the first and the last images of the stack, as the intensity level in the images can change during an acquisition. As we thus obtain two different threshold levels, we smooth the threshold linearly throughout the stack of images. 

The second step is to facilitate the calculation over the scratch. The user is asked to draw a line along the scratch, helping to evaluate the direction of the scratch. The images are then rotated to have the scratch placed vertically on the images.

We then remove any remaining pixels which do not belong to the scratch by drawing an approximate contour of the wound, which is applied to all of the images. If the scratch is closing, the contouring is done on the first image (as it is on this image that the scratch is the largest). For an expanding scratch, the contouring is done on the last image. The images are also resized to crop the edges of the images before rotation. 

The stack of images is finally ready for calculation.
First, we calculate wound closure \cite{grada_research_2017}, which evaluates the evolution of the area of the scratch over the acquisition period. The area is calculated on each frame by counting the pixels contained within the region of the scratch. The formula of the wound closure is: $\frac{area(t=0) - area(t)}{area(t=0)}\times 100$, giving a percentage of closure or expansion.

Secondly, we calculate wound size evolution. The width of the scratch is calculated by counting the pixels in each row of the image (as the scratch is oriented vertically, the width corresponds to an horizontal line, i.e. a row). Then, for each frame, we calculate the maximum and the minimum widths, but also the mean width and the 50$^{th}$ percentile (i.e. the median). The mean width and the median width were plotted over time and fitted with a bi-exponential function, determining the characteristic time and the speed of closing. The limit of closing was fixed at around 1-2 $\mu$m, which is the average interstice between cells far from the wound.

We validated our custom developed SAVE profiler software against an existing software. Cell Profiler is a cell image analysis software developed in 2005 \cite{lamprecht_cellprofiler_2007}. We used the Wound Healing example available on the website \url{https://cellprofiler.org}. By analysing images of a healing wound over time, this software calculates the area occupied by the cells on each image. Thus, we were able to calculate the wound closure (formula explained above) in order to compare with the results obtained with our SAVE profiler. We observed that Cell profiler was efficient on non closing wounds (i.e. failure of closing and expansion) and gives similar results. However, the segmentation performed by Cell profiler on closing wounds was incomplete (as shown in Fig.\ref{fig_1}~M and O) as it takes into account the differences of intensity in the images, misrepresenting the results of wound closure. Moreover, calculation time using Cell Profiler is considerably extended for long acquisitions (e.g. twice as long for more than 150 images), compared to our SAVE profiler.

\subsection*{D-FFOCT and immunochemistry comparison}

D-FFOCT has already been shown as a way to distinguish different states of a cell. In \cite{scholler_probing_2019}, it was shown that, while comparing FFOCT and D-FFOCT images of cell cultures, we can differentiate living cells from dying or dead cells, as these cells show a different dynamic profile. In \cite{scholler_dynamic_2020}, we showed that dead cells can be identified from D-FFOCT images alone. Also, different cell types (RPE, photoreceptors, inner retinal neurons) were shown to be distinguishable with D-FFOCT alone. 

In this article, we wished to identify the specific organelles responsible for D-FFOCT signal generation, and hence compared to immunohistochemistry. 

\section*{Acknowledgements}

The authors would like to thank Valérie Forster for providing primary porcine samples. The authors thank Pedro Mecê and Olivier Thouvenin for fruitful discussions on the results. The authors thank Marie Darche and Leyna Boucherit for their help in immunohistochemistry experiments and analysis of results. We would like to thank the direction and management teams of the institutions involved.
The authors thank the following sources of funding: OREO [ANR-19-CE19-0023], IHU FOReSIGHT [ANR-18-IAHU-0001], HELMHOLTZ (European Research Council (ERC) (\#610110), OPTORETINA  (European Research Council (ERC) (\#101001841), LabEx LIFESENSES [ANR-10-LABX-65], Institut Carnot Fondation Voir et Entendre, RETINIT-iPS [ANR-19-CE18-005].

\bibliographystyle{unsrt}
\bibliography{main.bib}

\end{document}